\documentclass[conference,10pt]{IEEEtran}
\IEEEoverridecommandlockouts

\usepackage{epsfig,rotating,setspace,latexsym,amsmath,epsf,amssymb,amsfonts,bm,theorem,cite,algorithm,graphicx,epsf,authblk,epstopdf,color,algpseudocode,bbm}

\newtheorem{theorem}{Theorem}
\newtheorem{lemma}{Lemma}
\newenvironment{Proof}[1]{\medskip\par\noindent{\bf Proof:\,}\,#1}{{\mbox{\,$\blacksquare$}\par}}

\allowdisplaybreaks

\begin{document}

\title{Age-Minimal Online Policies for Energy Harvesting Sensors with Random Battery Recharges\thanks{This research was supported in part by the National Science Foundation under Grants ECCS-1650299, CCF 14-22111, and CNS 15-26608.}}

\author[1]{Ahmed Arafa}
\author[2]{Jing Yang}
\author[3]{Sennur Ulukus}
\affil[1]{\normalsize Electrical Engineering Department, Princeton University}
\affil[2]{\normalsize School of Electrical Engineering and Computer Science, Pennsylvania State University}
\affil[3]{\normalsize Department of Electrical and Computer Engineering, University of Maryland}

\maketitle

\begin{abstract}
We consider an energy harvesting sensor that is sending measurement updates regarding some physical phenomenon to a destination. The sensor relies on energy harvested from nature to measure and send its updates, and is equipped with a battery of finite size to collect its harvested energy. The energy harvesting process is Poisson with unit rate, and arrives in amounts that {\it fully} recharge the battery. Our setting is online in the sense that the times of energy arrivals are revealed causally to the sensor after the energy is harvested; only the statistics of the arrival process is known a priori. Updates need to be sent in a timely manner to the destination, namely, such that the long term average {\it age of information} is minimized over the course of communication. The age of information is defined as the time elapsed since the freshest update has reached the destination. We first show that the optimal scheduling update policy is a {\it renewal policy,} and then show that it has a {\it multi threshold} structure: the sensor sends an update only if the age of information grows above a certain threshold that depends on the available energy.
\end{abstract}

\section{Introduction}

An energy harvesting sensor is sending measurement updates from a physical phenomenon to a destination. The sensor relies on energy harvested from nature to measure and send its updates, and has a {\it finite-sized} battery to save its incoming energy. The setting is {\it online}, and updates are to be sent such that the long term average age of information is minimized. Age of information is defined as the time elapsed since the freshest update has reached the destination.

Energy management and control in energy harvesting communication systems has been extensively studied in the literature for both offline and online settings. Earlier offline works study single-user channels \cite{jingP2P, kayaEmax, omurFade, ruiZhangEH}; multiuser channels \cite{jingBC, jingMAC}; and two-hop, relay, and two-way channels \cite{ruiZhangRelay, berkDiamond-jour, varan_twc_jour, arafa_baknina_twc_dec_proc}. Recent works in the online literature include the near-optimal results for single-user and multiuser channels \cite{ozgur_online_su, ozgur_online_mac, baknina_online_mac, baknina_online_bc}, systems with processing costs \cite{baknina-online-proc}, and systems with general utilities \cite{arafa-baknina-gnrl-online}.


Minimizing the age of information has been considered through mainly a queuing-theoretic framework, with a single source \cite{yates_age_1}; multiple sources \cite{yates_age_mac}; variations of the single source system such as randomly arriving updates \cite{ephremides_age_random}, update management and control \cite{ephremides_age_management}, and nonlinear age metrics \cite{ ephremides_age_non_linear}; multi hop networks \cite{shroff_age_multi_hop}; and caching systems \cite{yates-age-cache}.

Our work is most closely related to \cite{yates_age_eh, elif_age_eh, arafa-age-2hop, arafa-age-var-serv, elif-age-Emax, jing-age-online}, where age minimization in energy harvesting channels is considered. With the exception of \cite{arafa-age-var-serv}, an underlying assumption in these works is that energy expenditure is normalized; it takes one energy unit to measure and send an update to the destination. References  \cite{yates_age_eh, elif_age_eh} study a system with infinite size battery, with \cite{yates_age_eh} considering online scheduling with random service times (time for the update to take effect), and \cite{elif_age_eh} considering offline and online scheduling with zero service times. The offline policy in \cite{elif_age_eh} is extended to fixed non-zero service times in \cite{arafa-age-2hop} for single and multi hop settings, and to energy-controlled service times in \cite{arafa-age-var-serv}. The online policy in \cite{elif_age_eh} is found by dynamic programming in a discrete-time setting, and was shown to be of a threshold structure, where an update is sent only if the age of information is higher than a certain threshold. Motivated by the results in the infinite battery case, \cite{elif-age-Emax} then analyzes the performance of threshold policies under finite size battery and varying channel assumptions, yet with no proof of optimality. Recently, reference \cite{jing-age-online} proved the optimality of threshold policies when the battery size is equal to one unit using tools from renewal theory, and also provided an update policy that is asymptotically optimal when the battery size grows infinitely large. 

In this work, we extend the results of \cite{jing-age-online} and formally prove the optimality of online threshold policies for any finite-sized battery that gets randomly {\it fully} recharged over time according to a Poisson energy harvesting process with unit rate. That is, whenever energy arrives, it completely fills up the battery of the sensor. The goal is to optimally choose online feasible update transmission times such that the long term average age of information is minimized. We first show that the optimal update policy is a renewal policy. Then, we show that it has a {\it multi threshold} structure: the sensor sends a new update only if the age grows above a certain threshold that depends on the amount of energy in its battery.

\section{System Model and Problem Formulation}

We consider a sensor node that collects measurements from a physical phenomenon and sends updates to a destination over time. The sensor relies on energy harvested from nature to acquire and send its measurement updates, and is equipped with a battery of finite size $B$ to save its incoming energy. The sensor consumes one unit of energy to measure and send out an update to the destination. We assume that updates are sent over an error-free link with negligible transmission times as in \cite{elif_age_eh, elif-age-Emax, jing-age-online}. Energy arrives (is harvested) in $B$ units at times $\{t_1,t_2,\dots\}$ according to a Poisson process of rate $1$. This models, e.g., situations where the battery size is relatively small with respect to the amounts of harvested energy, and hence energy arrivals fully recharge the battery. We note that this random battery recharging model has been previously considered in the online scheduling literature in \cite{ozgur_online_su, ozgur_online_mac, baknina_online_mac, baknina_online_bc, baknina-online-proc, arafa-baknina-gnrl-online} and in the information-theoretic approach considered in \cite{ozgur_rbr}. Our setting is online in which energy arrival times are revealed causally over time; only the arrival rate is known a priori.

Let $s_i$ denote the time at which the sensor acquires (and transmits) the $i$th measurement update, and let $\mathcal{E}(t)$ denote the amount of energy remaining in the battery at time $t$. We then have the following energy causality constraint \cite{jingP2P}
\begin{align} \label{eq_en_caus}
\mathcal{E}\left(s_i^-\right)\geq1,\quad\forall i
\end{align}
We assume that we begin with a full battery at time $0$, and that the battery evolves as follows over time
\begin{align} \label{eq_battery}
\mathcal{E}\left(s_i^-\right)=\min\left\{\mathcal{E}\left(s_{i-1}^-\right)-1+ B\cdot\mathcal{A}\left(x_i\right),B\right\}
\end{align}
where $x_i\triangleq s_i-s_{i-1}$, and $\mathcal{A}(x_i)$ denotes the number of energy arrivals in $[s_{i-1},s_i)$. Note that $\mathcal{A}(x_i)$ is a Poisson random variable with parameter $x_i$. We denote by $\mathcal{F}$ the set of feasible transmission times $\{s_i\}$ described by (\ref{eq_en_caus}) and (\ref{eq_battery}) in addition to a full battery at time 0, i.e., $\mathcal{E}(0)=B$.

The goal is to choose an online feasible transmission policy $\{s_i\}$ (or equivalently $\{x_i\}$) such that the long term average of the age of information experienced at the destination is minimized. The age of information is defined as the time elapsed since the latest update has reached the destination. The age at time $t$ is formally defined as
\begin{align}
a(t)\triangleq t-u(t)
\end{align}
where $u(t)$ is the time stamp of the latest update received before time $t$. Let $n(t)$ denote the total number of updates sent by time $t$. We are interested in minimizing area under the age curve. At time $t$, this is given by
\begin{align} \label{eq_aoi}
r(t)\triangleq\frac{1}{2}\sum_{i=1}^{n(t)}x_i^2+\frac{1}{2}\left(t-s_{n(t)}\right)^2
\end{align}
and therefore the goal is to characterize the following quantity
\begin{align} \label{opt_main}
\bar{r}\triangleq\min_{{\bm x}\in\mathcal{F}}\limsup_{T\rightarrow\infty}\frac{1}{T}\mathbb{E}\left[r(T)\right]
\end{align}
where $\mathbb{E}(\cdot)$ is the expectation operator.

\section{Unit-Sized Battery}

In this section, we review the case $B=1$ studied in \cite{jing-age-online}, where the authors first show that renewal policies, i.e., policies with update times $\{s_i\}$ forming a renewal process, outperform any other {\it uniformly bounded policy}, which are policies whose inter-update delays, as functions of the energy inter-arrival times, have a bounded second moment \cite[Definition 3]{jing-age-online}. Then, it is shown that the optimal renewal policy is a threshold policy, where an update is sent only if the age of information grows above a certain threshold. We review this latter result in this section.

Let $\tau_i$ denote the time until the next energy arrival since the $i-1$st update time, $s_{i-1}$. Since the arrival process is Poisson with rate 1, $\tau_i$'s are independent and identically distributed (i.i.d.) exponential random variables with parameter 1. Under renewal policies, the $i$th inter-update time $x_i$ should not depend on the events before $s_{i-1}$; it can only be a function of $\tau_i$. Moreover, under any feasible policy, $x_i(\tau_i)$ cannot be smaller than $\tau_i$, since the battery is empty at $s_{i-1}$. Next, note that whenever an update occurs, both the battery and the age drop to 0, and hence the system resets. This constitutes a renewal event, and therefore using the laws of large number of renewal processes \cite{ross_stochastic}, problem (\ref{opt_main}) reduces to
\begin{align} \label{opt_b1}
\bar{r}=\min_{x(\tau)\geq\tau}\quad&\frac{\mathbb{E}\left[x(\tau)^2\right]}{2\mathbb{E}[x(\tau)]}
\end{align}
where expectation is over the exponential random variable $\tau$.

In order to make problem (\ref{opt_b1}) more tractable to solve, we introduce the following parameterized problem
\begin{align} \label{opt_b1_lmda}
p_1(\lambda)\triangleq\min_{x(\tau)\geq\tau}\quad&\frac{1}{2}\mathbb{E}\left[x(\tau)^2\right]-\lambda\mathbb{E}[x(\tau)]
\end{align}
This approach has also been used in \cite{sun-weiner}. One can show that $p_1(\lambda)$ is decreasing in $\lambda$, and that the optimal solution of problem (\ref{opt_b1}) is given by $\lambda^*$ satisfying $p_1(\lambda^*)=0$. Focusing on problem (\ref{opt_b1_lmda}), we introduce the following Lagrangian \cite{boyd}
\begin{align}
\mathcal{L}=&\frac{1}{2}\int_0^\infty x^2(\tau)e^{-\tau}d\tau-\lambda\int_0^\infty x(\tau)e^{-\tau}d\tau \nonumber\\
&-\int_0^\infty\mu(\tau)\left(x(\tau)-\tau\right)d\tau
\end{align}
where $\mu(\tau)$ is a non-negative Lagrange multiplier. Taking derivative with respect to $x(t)$ and equating to 0 we get
\begin{align} \label{eq_x_b1_kkt}
x(t)=\lambda+\frac{\mu(t)}{e^{-t}}
\end{align}
We now have two cases: 1) $t\leq\lambda:$ in this case we cannot have $x(t)=t$, or else the left hand side of (\ref{eq_x_b1_kkt}) would be smaller than the right hand side; 2) $t>\lambda:$ in this case, we cannot have $x(t)>t$, or else by complementary slackness \cite{boyd} $\mu(t)=0$ and the left hand side of (\ref{eq_x_b1_kkt}) would be larger than the right hand side. We conclude that the optimal $x(t)$ is given by
\begin{align}
x(t)=\begin{cases}\lambda,\quad &t\leq\lambda\\
t,\quad &t>\lambda\end{cases}
\end{align}

This means that the optimal inter-update time is threshold-based; if an energy arrival occurs before $\lambda$ amount of time since the last update time, i.e., if $\tau<\lambda$, then the sensor should not use this energy amount right away to send an update. Instead, it should wait for $\lambda-\tau$ extra amount of time before updating. Else, if an energy arrival occurs after $\lambda$ amount of time since the last update time, i.e., if $\tau\geq\lambda$, then the sensor uses that amount of energy to send an update right away. We coin this kind of policy {\it $\lambda$-threshold policy}. Substituting this $x(t)$ into problem (\ref{opt_b1_lmda}) we get
\begin{align} \label{eq_p1_lmda}
p_1(\lambda)=e^{-\lambda}-\frac{1}{2}\lambda^2
\end{align}
which admits a unique solution of $\lambda^*\approx0.9012$ when equated to 0. Next, we extend the approach in this section to characterize optimal policies for larger (general) battery sizes.

\section{The General Case}

In this section, we focus on the case of $B=k$ energy units for some positive integer $k\geq2$. Let $l_i$ denote the $i$th time that the battery level falls down to $k-1$ energy units. We use the term {\it epoch} to denote the time duration between two consecutive such events, and define $x_{k,i}\triangleq l_i-l_{i-1}$ as the length of the $i$th epoch. The main reason behind choosing such specific event to determine the epoch's start/end times is that the epoch would then contain at most $k$ updates, and that any other choice leads to having possibly infinite number of updates in a single epoch, which is clearly more complex to analyze. Let $\tau_i$ denote the time until the next energy arrival after $l_{i-1}$.  One scenario for the update process in the $i$th epoch would be that starting at time $l_{i-1}$, the sensor sends an update only after the battery recharges, i.e., at some time after $l_{i-1}+\tau_i$, causing the battery state to fall down from $k$ to $k-1$ again. Another scenario would be that the sensor sends $j\leq k-1$ updates before the battery recharges, i.e., at some times before $l_{i-1}+\tau_i$, and then submits one more update after the recharge occurs, making in total $j+1$ updates in the $i$th epoch.

\begin{figure}[t]
\center
\includegraphics[scale=.9]{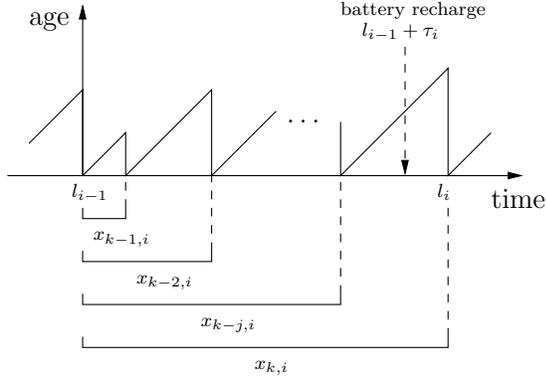}
\caption{Age evolution over time in the $i$th epoch, with $j+1\leq k$ updates.}
\label{fig_age_ex}
\vspace{-.2in}
\end{figure}

Let us now define $x_{j,i}$, $1\leq j\leq k-1$, to be the time it takes the sensor to send $k-j$ updates in the $i$th epoch before a battery recharge occurs. That is, starting at time $l_{i-1}$, and assuming that the $i$th epoch contains $k$ updates, the sensor sends the first update at $l_{i-1}+x_{k-1,i}$, followed by the second update at $l_{i-1}+x_{k-2,i}$, and so on, until it submits the $k-1$st update at $l_{i-1}+x_{1,i}$, using up all the energy in its battery. The sensor then waits until it gets a recharge at $l_{i-1}+\tau_i$ before sending its final $k$th update in the epoch. See Fig.~\ref{fig_age_ex} for an example run of the age of information curve during the $i$th epoch given that the sensor sends $j+1\leq k$ updates.

In general, under any feasible status updating online policy, $\{x_{j,i}\}_{j=1}^{k-1}$ and $x_{k,i}$ may depend on all the history of status updating and energy arrival information up to $l_{i-1}$, which we denote by $\mathcal{H}_{i-1}$. In addition to that, the value of $x_{k,i}$ can also depend on $\tau_i$. However, by the energy causality constraint (\ref{eq_en_caus}), the values of $\{x_{j,i}\}_{j=1}^{k-1}$ cannot depend on $\tau_i$. This is due to the fact that if the sensor updates $j+1$ times in the same epoch, then the first $j$ updates should occur before the battery recharges. Focusing on uniformly bounded policies, we now have the following theorem.

\begin{figure*}[t]
\begin{align}
\frac{\mathbb{E}\left[\sum_{i=1}^\infty R\left({\bm x}_i\right)\mathbbm{1}_{i\leq N_T}\right]}{\mathbb{E}\left[\sum_{i=1}^\infty x_{k,i}\mathbbm{1}_{i\leq N_T}\right]} &=\frac{\sum_{i=1}^\infty\mathbb{E}_{\mathcal{H}_{i-1}}\left[\mathbb{E}_{\tau_i}\left[\hat{R}_i\left(\gamma,\mathcal{H}_{i-1}\right)\right]\cdot\mathbbm{1}_{i\leq N_T}\Big|\mathcal{H}_{i-1}\right]}{\mathbb{E}\left[\sum_{i=1}^\infty x_{k,i}\mathbbm{1}_{i\leq N_T}\right]} \label{eq_pf_ren_2} \\
&=\frac{\sum_{i=1}^\infty\mathbb{E}_{\mathcal{H}_{i-1}}\left[\mathbb{E}_{\tau_i}\left[\hat{x}_{k,i}\left(\gamma,\mathcal{H}_{i-1}\right)\right]
\cdot \frac{\mathbb{E}_{\tau_i}\left[\hat{R}_i\left(\gamma,\mathcal{H}_{i-1}\right)\right]}{\mathbb{E}_{\tau_i}\left[\hat{x}_{k,i}\left(\gamma,\mathcal{H}_{i-1}\right)\right]} 
\cdot \mathbbm{1}_{i\leq N_T}\Big|\mathcal{H}_{i-1}\right]}{\mathbb{E}\left[\sum_{i=1}^\infty x_{k,i}\mathbbm{1}_{i\leq N_T}\right]} \\
&\geq\frac{\sum_{i=1}^\infty\mathbb{E}_{\mathcal{H}_{i-1}}\left[\mathbb{E}_{\tau_i}\left[\hat{x}_{k,i}\left(\gamma,\mathcal{H}_{i-1}\right)\right]
\cdot R^*\left(\mathcal{H}_{i-1}\right)
\cdot \mathbbm{1}_{i\leq N_T}\Big|\mathcal{H}_{i-1}\right]}{\mathbb{E}\left[\sum_{i=1}^\infty x_{k,i}\mathbbm{1}_{i\leq N_T}\right]}\geq R_{\min}
\end{align}
\hrulefill
\vspace{-.2in}
\end{figure*}

\begin{theorem} \label{thm_ren}
The optimal status update policy for the case $B=k$ is a renewal policy, i.e., the sequence $\{l_i\}$ forms a renewal process. Moreover, the optimal $\{x_{j,i}\}_{j=1}^{k-1}$ are constants, and the optimal $x_{k,i}$ only depends on $\tau_i$.
\end{theorem}

\begin{Proof}
Consider any feasible uniformly bounded policy. Let ${\bm x}_i\triangleq\{x_{1,i},\dots,x_{k,i}\}$, and let us denote by $R\left({\bm x}_i\right)$ the area under the age curve during the $i$th epoch. Then
\begin{align}
R\left({\bm x}_i\right)=&\frac{1}{2}\sum_{j=1}^{k-1}\left(x_{j,i}-x_{j+1,i}\right)^2\mathbbm{1}_{x_{j,i}\leq\tau_i} \nonumber \\
&+\frac{1}{2}\left(x_{k,i}\left(\tau_i\right)-\sum_{j=1}^{k-1}x_{j,i}\mathbbm{1}_{x_{j,i}\leq\tau_i<x_{j-1,i}}\right)^2
\end{align} 
where $\mathbbm{1}_A$ equals 1 if the event $A$ is true, and 0 otherwise. Next, for a given time $T$, let $N_T$ denote the number of epochs that have already {\it started} by time $T$, and for a fixed history $\mathcal{H}_{i-1}$, let us group all the status updating sample paths that have the same $\tau_i$ and perform a statistical averaging over all of them to get the following average age in the $i$th epoch
\begin{align}
\hat{R}_i\left(\gamma,\mathcal{H}_{i-1}\right)\triangleq\mathbb{E}\left[R\left({\bm x}_i\right)|\tau_i=\gamma,\mathcal{H}_{i-1}\right]
\end{align}
Then, we have
\begin{align}
\mathbb{E}&\left[R\left({\bm x}_i\right)\mathbbm{1}_{i\leq N_T}\right] \nonumber \\
&=\mathbb{E}_{\mathcal{H}_{i-1}}\left[\mathbb{E}_{\tau_i}\left[\hat{R}_i\left(\gamma,\mathcal{H}_{i-1}\right)\right]\cdot\mathbbm{1}_{i\leq N_T}\Big|\mathcal{H}_{i-1}\right] \label{eq_pf_ren_1}
\end{align}
where equality follows since $\mathbbm{1}_{i\leq N_T}$ is independent of $\tau_i$ given $\mathcal{H}_{i-1}$. Similarly, define the average $i$th epoch length as
\begin{align}
\hat{x}_{k,i}\left(\gamma,\mathcal{H}_{i-1}\right)\triangleq\mathbb{E}\left[x_{k,i}|\tau_i=\gamma,\mathcal{H}_{i-1}\right]
\end{align}

Next, note that by (\ref{eq_aoi}), the following holds
\begin{align} \label{eq_aoi_bd}
\frac{1}{T}\sum_{i=1}^\infty R_i\mathbbm{1}_{i\leq N_T-1}\leq \frac{r(T)}{T} \leq \frac{1}{T}\sum_{i=1}^\infty R_i\mathbbm{1}_{i\leq N_T}
\end{align} 
Following similar analysis as in \cite[Appendix C-1]{jing-age-online}, one can show that $\lim_{T\rightarrow\infty} \frac{\mathbb{E}\left[R_{N_T}\right]}{T}=0$
for any uniformly bounded policy. Hence, in the sequel, we derive a lower bound on $\frac{1}{T}\mathbb{E}\left[\sum_{i=1}^\infty R_i\mathbbm{1}_{i\leq N_T}\right]$ and use the above note to conclude that it is also a lower bound on $\frac{\mathbb{E}\left[r(T)\right]}{T}$ as $T\rightarrow\infty$. Towards that end, note that $\mathbb{E}\left[\sum_{i=1}^\infty x_{k,i}\mathbbm{1}_{i\leq N_T}\right]\geq T$. Then, we have
\begin{align}
\frac{1}{T}\mathbb{E}\left[\sum_{i=1}^\infty R\left({\bm x}_i\right)\mathbbm{1}_{i\leq N_T}\right] \geq \frac{\mathbb{E}\left[\sum_{i=1}^\infty R\left({\bm x}_i\right)\mathbbm{1}_{i\leq N_T}\right]}{\mathbb{E}\left[\sum_{i=1}^\infty x_{k,i}\mathbbm{1}_{i\leq N_T}\right]}
\end{align}
Next, we proceed by lower bounding the right hand side of the above equation through a series of equations at the top of this page. In there, (\ref{eq_pf_ren_2}) follows from (\ref{eq_pf_ren_1}) and the monotone convergence theorem, $R^*\left(\mathcal{H}_{i-1}\right)$ is the minimum value of $\frac{\mathbb{E}_{\tau_i}\left[\hat{R}_i\left(\gamma,\mathcal{H}_{i-1}\right)\right]}{\mathbb{E}_{\tau_i}\left[\hat{x}_{k,i}\left(\gamma,\mathcal{H}_{i-1}\right)\right]}$, and $R_{\min}$ is the minimum value of $R^*\left(\mathcal{H}_{i-1}\right)$ over all possible epochs and their corresponding histories, i.e., the minimum over all $i$ and $\mathcal{H}_{i-1}$.

Observe that a policy achieving $R^*\left(\mathcal{H}_{i-1}\right)$ is a policy where $\{x_{j,i}\}_{j=1}^{k-1}$ are constants and $x_{k,i}$ is a function of $\tau_i$ only, since the history $\mathcal{H}_{i-1}$ is fixed. Now, if we repeat the policy that achieves $R_{\min}$ over all epochs, we get a renewal policy where $\forall i$ $\{x_{j,i}\}_{j=1}^{k-1}$ are constants and $x_{k,i}$ is only a function of $\tau_i$. Since $\tau_i$'s are i.i.d., the epoch lengths are also i.i.d., and $\{l_i\}$ forms a renewal process. This completes the proof.
\end{Proof}

Theorem~\ref{thm_ren} indicates that the sensor should let its battery fall down to $k-1$ at times that constitute a renewal policy. Next, we characterize the optimal renewal policy by which the sensor sends its updates. Using the strong law of large numbers of renewal processes \cite{ross_stochastic}, problem (\ref{opt_main}) reduces to
\begin{align} \label{opt_bk}
\bar{r}=\min_{{\bm x}}\quad&\frac{\mathbb{E}\left[R\left({\bm x}\right)\right]}{\mathbb{E}\left[x_k(\tau)\right]} \nonumber \\
\mbox{s.t.}\quad&x_{k-1}\geq0 \nonumber\\
&x_{j-1}\geq x_j,\quad 2\leq j\leq k-1 \nonumber \\
&x_k(\tau)\geq\tau,\quad\forall\tau
\end{align}
where the expectation is over the exponential random variable $\tau$. Similar to the $B=1$ case, we define $p_k(\lambda)$ as follows
\begin{align} \label{opt_bk_lmda}
p_k(\lambda) \triangleq \min_{{\bm x}} \quad &\mathbb{E}\left[R\left({\bm x}\right)\right]-\lambda\mathbb{E}\left[x_k(\tau)\right] \nonumber \\
\mbox{s.t.} \quad &\text{constraints of (\ref{opt_bk})}
\end{align}
As in the $B=1$ case, one can show that $p_k(\lambda)$ is decreasing in $\lambda$, and the optimal solution of problem (\ref{opt_bk}) is given by $\lambda^*$ satisfying $p_k(\lambda^*)=0$. We omit the details due to space limits.

\begin{figure}[t]
\center
\includegraphics[scale=.85]{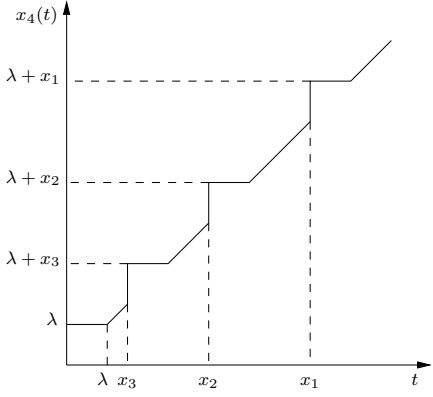}
\caption{Optimal multi threshold structure of $x_4(t)$.}
\label{fig_th_policy}
\vspace{-.2in}
\end{figure}

Since the optimal solution for the $B=k$ case cannot be larger than that of the $B=1$ case, which is $0.9012$, one can use, e.g., a bisection search over $(0,0.9012]$ to find the optimal $\lambda$ for $B=k$. We now write the following Lagrangian for problem (\ref{opt_bk_lmda}) after expanding the objective function
\begin{align}
\mathcal{L}=&\frac{1}{2}x_{k-1}^2e^{-x_{k-1}}+\frac{1}{2}\sum_{j=1}^{k-2}\left(x_j-x_{j+1}\right)^2e^{-x_j} \nonumber\\
&+\frac{1}{2}\int_0^{x_{k-1}}\!\!\!x_k(\tau)^2e^{-\tau}d\tau \nonumber\\
&+\frac{1}{2}\sum_{j=2}^{k-1} \! \int_{x_j}^{x_{j-1}} \!\!\! \left(x_k(\tau)-x_j\right)^2e^{-\tau}d\tau \! - \! \lambda \! \int_0^\infty \!\!\! x_k(\tau)e^{-\tau}d\tau \nonumber\\
&+\frac{1}{2}\int_{x_1}^\infty\left(x_k(\tau)-x_1\right)^2e^{-\tau}d\tau -\mu_{k-1}x_{k-1} \nonumber\\
&-\sum_{j=1}^{k-2}\mu_j\left(x_j-x_{j+1}\right) -\int_0^\infty\!\!\!\mu_k(\tau)\left(x_k(\tau)-\tau\right)d\tau
\end{align}
where $\left\{\mu_1,\dots,\mu_{k-1},\mu_k(\tau)\right\}$ are non-negative Lagrange multipliers. Taking derivative with respect to $x_k(t)$ and equating to 0 we get
\begin{align}
x_k(t)=\lambda+\sum_{j=1}^{k-1}x_j\mathbbm{1}_{x_j\leq t<x_{j-1}}+\frac{\mu_k(t)}{e^{-t}}
\end{align}
Now let us assume that $\lambda$ is smaller than $\min\left\{x_{k-1},\min_{1\leq j\leq k-2}x_j-x_{j+1}\right\}$, and verify this assumption later on. Proceeding similarly to the analysis of the $B=1$ case, we get that
\begin{align} \label{eq_xk_opt}
x_k(t)=\begin{cases}\lambda,\quad&t<\lambda\\t,\quad&\lambda\leq t<x_{k-1}\\\lambda+x_{k-1},\quad&x_{k-1}\leq t<\lambda+x_{k-1}\\t,\quad&\lambda+x_{k-1}\leq t<x_{k-2}\\
\vdots\\
\lambda+x_1,\quad&x_1\leq t<\lambda+x_1\\
t,\quad&t\geq\lambda+x_1\end{cases}
\end{align}
A depiction of the above policy for $k=4$ is shown in Fig.~\ref{fig_th_policy}.

Thus, the optimal update policy has the following structure. Starting with a battery of $k-1$ energy units and zero age, if the next battery recharge occurs at any time before $\lambda$ time units, then the sensor updates at exactly $t=\lambda$. While if it occurs at any time between $\lambda$ and $x_{k-1}$, then the sensor updates right away. This is the same as the {\it $\lambda$-threshold policy}, the solution of the $B=1$ case, except that it has a cut-off at $t=x_{k-1}$. This cut-off value has the following interpretation: if the battery recharge does not occur until $t=x_{k-1}$, then the sensor updates at $t=x_{k-1}$, causing the battery and the age to fall down to $k-2$ and $0$, respectively. The sensor then repeats the {\it $\lambda$-threshold policy} described above with a new cut-off value of $x_{k-2}$, i.e., if the recharge does not occur until $t=x_{k-2}$, then the sensor updates again at $t=x_{k-2}$, causing the battery and the age to fall down to $k-3$ and $0$, respectively. This technique repeats up to $t=x_1$, when the sensor updates for the $k-1$st time, emptying its battery. At this time, the sensor waits for the battery recharge and applies the {\it $\lambda$-threshold policy} one last time, with no cut-off value, to submit the last $k$th update in the epoch. Note that if the battery recharge occurs at some time $\tau<x_{k-1}$, then there would be $1$ update in the epoch. On the other hand, if $x_j\leq\tau<x_{j-1}$, for some $2\leq j\leq k-1$, then there would be $k-j+1$ updates. Finally, if $\tau\geq x_1$ then there would be $k$ updates.

In the sequel we find the optimal values of $\{x_j\}_{j=1}^{k-1}$ (and $\lambda$) by taking derivatives of the Lagrangian with respect to $x_j$'s and equating to 0. Before doing so, we simplify the objective function of $p_k(\lambda)$ by evaluating the expectations involved and using (\ref{eq_xk_opt}). After some simplifications we get
\begin{align} \label{eq_obj_fn}
&\mathbb{E}\left[R\left({\bm x}\right)\right]-\lambda\mathbb{E}\left[x_k(\tau)\right]=e^{-\lambda}-\frac{1}{2}\lambda^2+f_1(\lambda)\sum_{j=1}^{k-1}e^{-x_j} \nonumber \\
&~~~-\left(x_{k-1}+1\right)e^{-x_{k-1}}-\sum_{j=1}^{k-2}\left(x_j-x_{j+1}+1\right)e^{-x_j}
\end{align}
where $f_1(\lambda)\triangleq\lambda+e^{-\lambda}-\frac{1}{2}\lambda^2$. Using the above in the Lagrangian and taking derivatives we get
\begin{align}
x_1&=x_2+f_1(\lambda)+\mu_1e^{x_1} \label{eq_kkt_x1} \\
x_j&=x_{j+1}+f_1(\lambda)+\left(\mu_j-\mu_{j-1}-e^{x_{j-1}}\right)e^{-x_j},\quad \nonumber \\ &\qquad\qquad\qquad\qquad\qquad\qquad\qquad2\leq j\leq k-2 \label{eq_kkt_xj} \\
x_{k-1}&=f_1(\lambda)+\left(\mu_{k-1}-\mu_{k-2}-e^{-x_{k-2}}\right)e^{x_{k-1}} \label{eq_kkt_xk1}
\end{align}
Now let us assume that $x_j>x_{j+1}$, $1\leq j\leq k-2$, and $x_{k-1}>0$. Hence, by complementary slackness we have $\mu_j=0$, $1\leq j\leq k-1$. One can then substitute $x_1-x_2$ in (\ref{eq_kkt_xj}) for $j=2$ to find $x_2-x_3$ and proceed recursively to get
\begin{align}
x_j-x_{j+1}&=f_j(\lambda),\quad 1\leq j\leq k-2 \label{eq_xj_lmda}\\
x_{k-1}&=f_{k-1}(\lambda) \label{eq_xk1_lmda}
\end{align}
where we have defined
\begin{align}
f_j(\lambda)&\triangleq f_1(\lambda)-e^{-f_{j-1}(\lambda)},\quad2\leq j\leq k-1
\end{align}
We have the following result on the structure of $\{f_j(\lambda)\}$; the proof follows by induction and is omitted due to space limits.
\begin{lemma} \label{thm_fj_dec}
For a fixed $\lambda$, the sequence $\{f_j(\lambda)\}_{j=1}^{k-1}$ is decreasing; and for a fixed $j$, $f_j(\lambda)$ is decreasing in $\lambda$.
\end{lemma}
%
%

Note that $f_j(\lambda)$ represents the inter-update delay between updates $k-j-1$ and the $k-j$. With this in mind, Lemma~\ref{thm_fj_dec} has an intuitive explanation; it shows that when the amount of energy in the battery is relatively low, the sensor becomes less eager to send the next update, so that it does not run out of energy. And oppositely when the amount of energy in the battery is relatively high, so that it makes use of the available energy before the next recharge overflows the battery. Next, by equations (\ref{eq_xj_lmda}) and (\ref{eq_xk1_lmda}), we proceed recursively from $j=k-1$ to $j=1$ to find the values of $x_j$'s in terms of $\lambda$. This gives
\begin{align}
x_j=\sum_{m=j}^{k-1}f_m(\lambda),\quad 1\leq j\leq k-1
\end{align}
Finally, we substitute the above in (\ref{eq_obj_fn}) to get 
\begin{align} \label{eq_pk_lmda}
p_k(\lambda)=&e^{-\lambda}-\frac{1}{2}\lambda^2+\sum_{j=1}^{k-1}\left(f_1(\lambda)-f_j(\lambda)-1\right)e^{-\sum_{m=j}^{k-1}f_j(\lambda)} \nonumber \\
=&e^{-\lambda}-\frac{1}{2}\lambda^2-e^{-f_{k-1}(\lambda)}
\end{align}
and perform a bisection search over $\lambda\in(0,0.9012]$ to find the optimal $\lambda^*$ that solves $p_k(\lambda^*)=0$. It is worth noting that for $k=1$, the summation in (\ref{eq_pk_lmda}) vanishes and we directly get (\ref{eq_p1_lmda}). Finally, observe that $p_k(\lambda)=0$ implies $f_{k-1}(\lambda)=-\log\left(e^{-\lambda}-\frac{1}{2}\lambda^2\right)$. Since $0<\lambda\leq0.9012$, then $0\leq e^{-\lambda}-\frac{1}{2}\lambda^2<1$, and hence $f_{k-1}(\lambda)>0$; moreover $f_{k-1}(\lambda)>-\log\left(e^{-\lambda}\right)=\lambda$. By Lemma~\ref{thm_fj_dec}, the above argument shows that: 1) $f_j\left(\lambda^*\right)>0$, $1\leq j\leq k-1$, which further implies by (\ref{eq_kkt_x1})-(\ref{eq_kkt_xk1}) that all Lagrange multipliers are zero, as previously assumed; 2) $\lambda^*<f_j\left(\lambda^*\right)$, $1\leq j\leq k-1$, which verifies the previous assumption regarding the optimal age being smaller than all inter-update delays.

\section{Numerical Examples}

We now discuss some numerical results. We compare the optimal policy derived in this work with two other update policies. The first is a best effort uniform updating policy, where the sensor aims at sending an update every $1/B$ time units whenever it has enough energy, and stays silent otherwise. The rationale is that since the energy arrivals are with unit rate and value $B$, then the effective arrival rate is $1/B$, by which the sensor aims at uniformly spreading its updates over time. The other policy is the battery aware adaptive update policy proposed in \cite{jing-age-online}, in which the sensor aims at sending its next update depending on the status of its battery. This policy was originally proposed for systems where energy arrives in one units, and not in $B$ units as in this work. Hence, we slightly modify by dividing the update rate by $B$: if the battery has more (resp. less) than $B/2$ units, the sensor aims at sending the next update after $1/B(1+\beta)$ (resp. $1/B(1-\beta)$) time units; and if the battery has exactly $B/2$ units, then the sensor aims at sending the next update after $1/B$ time units. We choose $\beta=\log(B)/B$ \cite{jing-age-online}. 

\begin{figure}[t]
\center
\includegraphics[scale=.45]{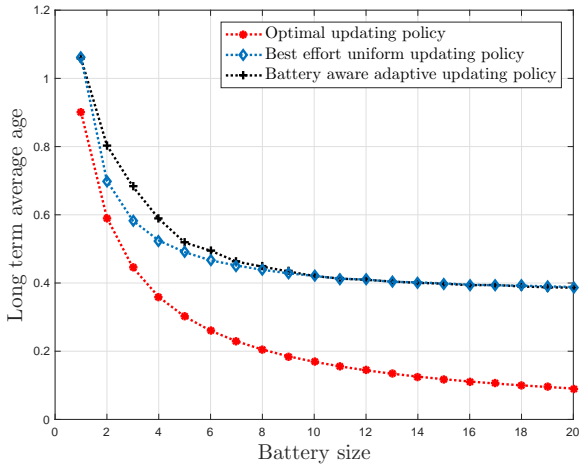}
\caption{Comparison of long term average age versus battery size under different update policies.}
\label{fig_age_battery}
\vspace{-.2in}
\end{figure}

In Fig.~\ref{fig_age_battery}, we plot the long term average age of the three policies versus battery size. We consider a system with $T=1000$ time units, and compute the long term average age over $1000$ iterations. We see from the figure that the optimal updating policy outperforms both uniform and battery aware adaptive updating policies, and that the gap between them grows larger with the battery size.

\section{Conclusions}

We characterized optimal online policies that minimize the long term average age of information in a single-user energy harvesting channel with a finite battery. Under Poisson energy arrivals that fully recharge the battery, we showed that the optimal update policy is a renewal policy, and that it has a multi threshold structure; the sensor sends a new update only after the age of information grows above a certain threshold that is a function of the available energy.

In our recent work \cite{arafa-age-sgl}, we complement the results in this paper and study online policies with {\it incremental} battery recharges where the energy arrives in single units, as opposed to full $B$ units in this work, and show that renewal-type policies are optimal. We also explicitly characterize the optimal renewal policy for the case $B=2$ and show that it has a threshold structure as well.


\end{document}